\documentclass[conference]{IEEEtran}
\IEEEoverridecommandlockouts
\usepackage{cite}
\usepackage{amsmath,amssymb,amsfonts}
\usepackage{algorithmic}
\usepackage{graphicx}
\usepackage{textcomp}
\usepackage{booktabs}
\usepackage{hyperref}
\usepackage{pifont}
\usepackage{url}
\usepackage{xcolor}
\usepackage[caption=false,font=footnotesize]{subfig}
\def\BibTeX{{\rm B\kern-.05em{\sc i\kern-.025em b}\kern-.08em
    T\kern-.1667em\lower.7ex\hbox{E}\kern-.125emX}}

\DeclareRobustCommand{\cmark}{\scalebox{1.9}{$\checkmark$}}

\DeclareRobustCommand{\xmark}{\scalebox{1.5}{\ding{55}}}

\begin{document}

\title{{\textsc{Argonaut}}: Interactive Visual Exploration for Distributed Optimization
}


\author{
\IEEEauthorblockN{
Srijoni Majumdar\IEEEauthorrefmark{1}, Chuhao Qin\IEEEauthorrefmark{1}, Evangelos Pournaras\IEEEauthorrefmark{1}\IEEEauthorrefmark{2}
}\
\IEEEauthorblockA{\IEEEauthorrefmark{1} School of Computer Science, University of Leeds, Leeds LS2 9JT, UK. E-mail: {S.Majumdar,C.Qin,E.Pournaras}@leeds.ac.uk}
\IEEEauthorblockA{\IEEEauthorrefmark{2} School of Energy Systems, LUT University, 53850 Lappeenranta, Finland}
}

\maketitle

\begin{abstract}
Distributed discrete-choice optimization in decentralized settings is often hard to explore and navigate: disentangling what other agents choose, how their choices are interdependent, and how they collectively reach a global objective quickly becomes intractable as the system scales. The major limitation is observability of the search process. Existing methods are largely centralized and offer limited support, visualizing only the final solution or providing algorithm backends over a fixed dataset, so how a solution is reached stays a black box. We present {\textsc{Argonaut}}, a lightweight, containerized optimization dashboard that enables interactive, visual exploration of the entire search process for multi-agent discrete-choice optimization in decentralized settings. Users upload datasets, construct agents and options, modify the decision space and its parameters on the fly, and run multiple algorithm backends to inspect how each configuration shapes local agent decisions and the resulting global objective. By uniting system construction, optimization, and analysis in one interactive loop, the first of its kind, {\textsc{Argonaut}} makes distributed discrete-choice optimization a human-in-the-loop process rather than a one-shot, black-box computation. We evaluate {\textsc{Argonaut}} on real-world household-electricity, shared-mobility, and sensor-data-exchange datasets scaling to 5600 agents and up to  1M solutions under brute force. Built on a Node.js interface with extensible Java and Python optimization backends, it maintains a typical runtime of $200$ agents over $100$ decision attributes in under $30$ seconds.
\end{abstract}


\section{Introduction}

Distributed discrete-choice optimization underpins several real-world problems, including scheduling, routing, and resource allocation. Each agent selects a preferred option from a finite set while collectively aiming to optimize a global objective. As agents and options scale, the decision space grows exponentially, making it increasingly difficult to work out how individual choices shape collective behavior. Decentralized settings make this harder still, as agents lack a complete view of one another's preferences and so of the decision space as a whole. The missing piece is the observability of the entire optimization process: how each agent chooses a plan, when that choice can change, how their choices are interdependent, and how they shape the global objective. This observability makes optimization practical and scalable for the real-time shared-resource decisions in energy, transport, mobility, logistics, and urban infrastructure~\cite{turan2024transition,pournaras2016self}. It leverages human oversight to steer the process through interventions, e.g., tuning parameters, restricting the decision space, or adjusting the global objective~\cite{xin2020general,blum2003metaheuristics,kotthoff2016algorithm}, helping agents align on power use, routes, or parking at  critical moments, and preventing the disruption that would otherwise follow.

\begin{figure}[h]
    \centering
    \includegraphics[width=0.42\textwidth]{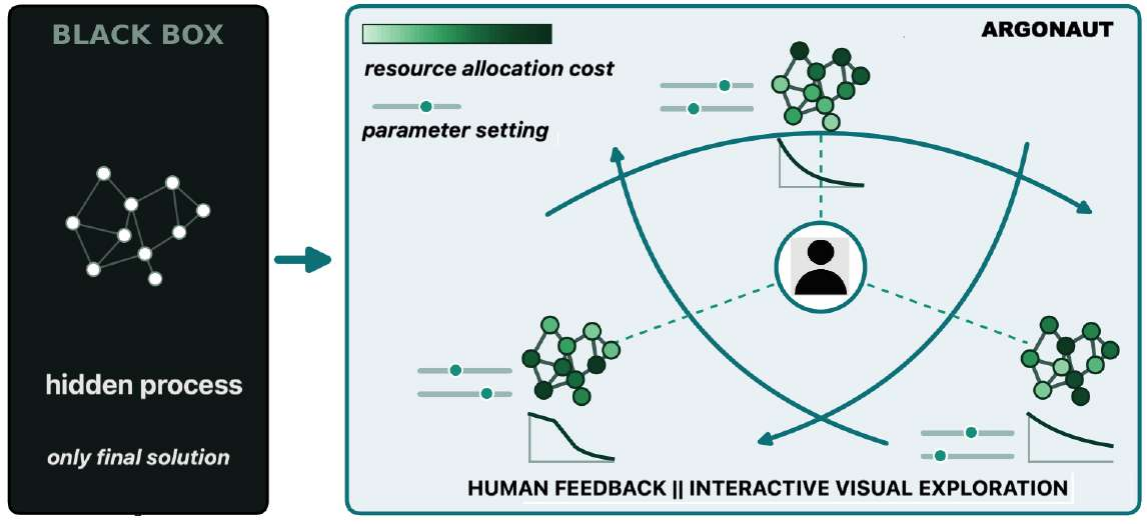}
    \caption{\textsc{Argonaut}: Black-Box optimization to visual interactive human in the loop exploration }
    \label{fig:black_box}
\end{figure}

Existing methods offer only a limited view of the process. Some tag whether a solution is a global optimum, letting users re-run the optimization with different parameters to escape a local one, but only for centralized algorithms over a small decision space~\cite{colella2020human,klau2010human}.  Others specifically visualize only the final solution or provide flexible execution through multiple algorithm backends, but only over a fixed set of datasets~\cite{iohanalyzer,ortools,gurobi}. The available support thus remains fragmented, lacking an end-to-end observability of the individual and collective agent behavior across the optimization process.

We introduce {\textsc{Argonaut}}\footnote{{\textsc{Argonaut}}—Cloud-hosted service at \url{https://argonautsim-382548405389.us-central1.run.app/} with video [https://youtu.be/NMoVSavMvd8] to run. Running on localhost: code and instructions at \url{https://github.com/TDI-Lab/Argonaut} for MacOS, Linux, and Windows. } to bridge this gap, bringing multiple discrete-choice optimization paradigms for decentralized settings into a single, lightweight, containerized dashboard. It supports \emph{brute-force} search over the full decision space; tree-structured decentralized iterative methods that coordinate agents to converge to a solution; and a combination of the two, which ranks that solution against the global optimum found by brute-force search. {\textsc{Argonaut}} enables \emph{human-in-the-loop optimization} by uniting construction, optimization, and analysis in one loop, thereby transforming the optimization process from a \emph{black box} into an observable, interactive, and visual one (Figure~\ref{fig:black_box}). Users upload datasets, construct agents and options, prune or edit the decision space on the fly, and run alternative algorithms over the same configuration, intervening to improve the optimality of the solution. We perform a cross-domain evaluation of {\textsc{Argonaut}}, running brute-force and decentralized search over real-world household-electricity, shared-mobility, and data-sharing datasets, alongside synthetic ones, ranging from $70$ to $5600$ agents with $64$ to $144$ decision scenarios and up to $1$M solutions under brute force. Over multiple runs, we test correctness, scalability, and effectiveness: the cloud-hosted service maintains a typical runtime (on localhost) of under $30$ seconds for $200$ agents over $100$ decision attributes on minimal computing resources, guided exploration improves solution optimality on average by $24.87\%$, and non-experts reach near-optimal solutions on par with domain experts.

This paper makes three contributions: \noindent  (i) The first platform of its kind to make the entire optimization process \emph{observable and steerable} in decentralized settings. (ii) A \emph{lightweight, portable, and customizable} design, a containerized runtime with algorithm-agnostic APIs to configure multiple backends, {\em deployable in the cloud or on standalone systems}. (iii) A \emph{fast and computationally efficient} engine, well suited for integration into standard transaction frameworks for  operational decisions such as load balancing and scheduling in city infrastructure.

The remainder of this paper is organized as follows. Section~\ref{sec:related}
reviews related work on 
optimization visualization. Section~\ref{sec:design} describes the design of
{\textsc{Argonaut}}, including decision-space construction,
optimization, and analysis. Section~\ref{sec:implementation} details the
implementation and deployment. Section~\ref{sec:evaluation} reports the
evaluation, and Section~\ref{sec:conclusions} concludes.

\section{Related Work} \label{sec:related}

Prior work on observability in discrete-choice optimization falls into four lines of work, each covering only part of the problem.
First, proposed methods tag each achieved solution only by whether it is a global optimum, without any ranking, letting users re-run the search to escape a local optimum; this supports human-guided search, but with only limited interactive control~\cite{colella2020human,klau2010human}. Second, algorithm-specific tools such as SwarmViz~\cite{jornod2015swarmviz} visually monitor particle swarm optimization methods~\cite{he2016improved,lee2019visualizing} and map high-dimensional Pareto fronts to lower dimensions for trade-off comparison, but remain bound to a particular algorithm, objective structure, or \emph{offline} setting.

Third, some methods use visual steering to configure machine-learning algorithms through hyperparameter tuning and design-space exploration~\cite{thole2020design,ramu2022survey,yadav2025handling}.

A fourth line of work covers general optimization software and solvers, including OR-Tools~\cite{ortools}, HeuristicLab~\cite{heuristiclab}, IOHanalyzer~\cite{iohanalyzer}, Gurobi~\cite{gurobi}, and branch-and-bound visualizers~\cite{visualizing_branch_and_bound}, as well as decentralized solvers such as COHDA~\cite{hinrichs2013cohda} and DCOP methods~\cite{yeoh2010bnb}. These bring multiple algorithm backends into one tool but mostly over fixed datasets, and return only final solutions without interactive inspection of the optimization process. Discrete-choice modeling
tools~\cite{apollo_choice_modelling,mariel2021software} mostly help with choice modeling and estimating model parameters from data rather than navigating the combinatorial
decision space of coordinated agents.

Across these tools, support is mostly applicable for centralized settings and lacks an end-to-end view, offering no account of how discrete choices are formed, pruned, traversed, and balanced across \emph{local} and \emph{global} objectives, which is critical for decentralized systems. {\textsc{Argonaut}} addresses this gap as lightweight, software
that supports the full \emph{construct--optimize--analyze} loop for
discrete-choice optimization. Table~\ref{tab:visualizer-comparison} summarizes
this comparison across seven capabilities spanning centralized and decentralized
settings.

\begin{table}[t]
\centering
\caption{Comparison of existing optimization visualizers and software against
{\textsc{Argonaut}}. \cmark{} indicates support, \xmark{} indicates no
support, and NA means not applicable.}
\label{tab:visualizer-comparison}
\resizebox{\columnwidth}{!}{%
\begin{tabular}{lccccc}
\toprule
\textbf{\Large Capability}
&
\textbf{
\shortstack{
\large Human-guided\\
\large search\\
\cite{colella2020human,klau2010human}
}}
&
\textbf{
\shortstack{
\large Algorithm/dataset-\\
\large specific\\
\cite{jornod2015swarmviz,he2016improved,lee2019visualizing}
}}
&
\textbf{
\shortstack{
\large Machine-learning\\
\large specific\\
\cite{thole2020design,ramu2022survey}
}}
&
\textbf{
\shortstack{
\large Standard solvers/\\
\large benchmarks\\
\cite{ortools,gurobi,iohanalyzer}
}}
&
\textbf{\textsc{\large Argonaut}}
\\
\midrule

{\Large Parameter modeling}
& \cmark
& \xmark
& \cmark
& \cmark
& \cmark
\\

{\Large Global-optimum detection}
& \cmark
& \xmark
& \cmark
& \xmark
& \cmark
\\

{\Large Visualization of final solution}
& \cmark
& \cmark
& \cmark
& \xmark
& \cmark
\\

{\Large Multiple algorithm backends}
& \cmark
& \xmark
& \xmark
& \cmark
& \cmark
\\

{\Large Local \& global analysis}
& \xmark
& \cmark
& \xmark
& \xmark
& \cmark
\\

{\Large Decentralized multi-agent support}
& \xmark
& \xmark
&  {\large NA}
& \xmark
& \cmark
\\

{\Large Construct--optimize--analyze loop}
& \xmark
& \xmark
& \xmark
& \xmark
& \cmark
\\

\bottomrule
\end{tabular}%
}
\end{table}

\section{Design of {\textsc{Argonaut}}} \label{sec:design} 

\subsection{Optimization Problem}

We model a typical socio-technical system as a set of \textit{agents}
$A=\{a_1,\ldots,a_n\}$, where each agent $a_i \in A$ is an autonomous
decision-making unit, such as a household, vehicle, shipment, station, or
user~\cite{pournaras2018decentralized}. Each agent $a_i$ holds a finite set of
\textit{plans} $P_i=\{p_{i1},\ldots,p_{ik}\}$, its feasible options, i.e., those
satisfying the agent's {\em local constraints}, such as capacity or availability.
Every plan describes what an agent would do across a set of \textit{decision
attributes}, such as how much power to draw at each time step, how to allocate a
resource, or what demand to place. Formally, a plan $p_{ij}\in P_i$ is a
$d$-dimensional vector, with one value per attribute. A \textit{collective
solution} $s=(p_1,\ldots,p_n)$ is then one choice of plan per agent, and the set
of all such combinations forms the \textit{decision space}
$\mathcal{S}=P_1 \times P_2 \times \cdots \times P_n$, of size
$|\mathcal{S}|=\prod_{i=1}^{n}|P_i|$.

\noindent Three objectives evaluate a collective solution. The \textit{global objective} measures system-level inefficiency through nonlinear measures such as residual sum of squares and root mean square error, which quantify how closely the collective outcome matches a \emph{target global objective bound}, or variance, which captures imbalance across agents, resources, or time~\cite{karunasingha2022root}. These are nonlinear because cost grows
disproportionately with imbalance: e.g., in an energy grid, a few households
drawing power at the same peak minute strains the network far more than the same
demand spread out, and in transport, vehicles converging on one road cause
congestion that rises steeply with load. The \textit{local objective} measures
per-agent discomfort, how far an agent's selected plan sits from its preferred
one, e.g., a household shifting its laundry to an off-peak hour, or a driver
taking a longer route to ease congestion. The \textit{unfairness objective}
measures how unevenly this discomfort is spread across agents, so that no single
household or driver bears all the inconvenience~\cite{western20099}.
{\textsc{Argonaut}} combines these into a single total cost $C$, weighted by
agent-controlled \textit{trade-off parameters} $\alpha,\beta,\gamma \ge 0$ with
$\alpha+\beta+\gamma=1$, which set their \emph{priority}. The goal is to
find a feasible collective solution that minimizes $C$.

{\textsc{Argonaut}} explores this decision space in two modes. (i) \noindent {\em Brute force search.}  It constructs the full decision space $\mathcal{S}$ as the Cartesian
product of every agent's plan set, evaluates the total cost $C$ of each
collective solution $s\in\mathcal{S}$, and returns the one with the lowest cost,
which is the global optimum. (ii) \noindent {\em Iterative decentralized search.}  The agents coordinate over successive iterations to refine their choices. In
each iteration, every agent $a_i$ selects the plan from its set $P_i$ that
minimizes the total cost $C$, given the current plans of the other agents. This
repeats until the objectives stabilize, converging to a near-optimal, or
the optimal solution. When
complete search is tractable, its global optimum serves as the reference against
which any optimality gap of the iterative solution is measured.

\begin{figure}[h]
    \centering
    \includegraphics[
        width=0.48\textwidth,
        height=0.32\textheight,
        keepaspectratio
    ]{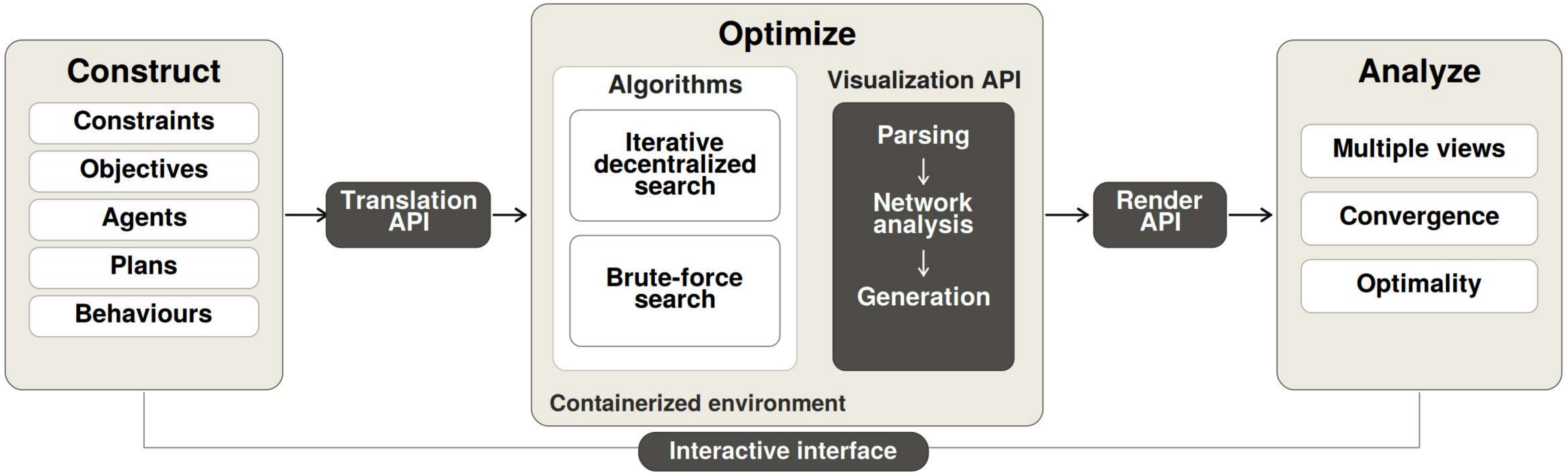}
    \caption{Schematic architectural view of {\textsc{Argonaut}}, showing the
    \emph{construct--optimize--analyze} loop from decision-space construction,
    through algorithmic optimization, to interactive visual analytics.}
    \label{fig:Design}
\end{figure}



\begin{figure}[t]
    \centering

    \subfloat[
        Construct component of {\textsc{Argonaut}}, Pre-configured dataset:
        \emph{$(A,\{P_i\},d)$}, Number of agents: \emph{$n=|A|$},
        Agent selection mode: \emph{subset $A'\subseteq A$}, Plans per agent:
        \emph{$|P_i|=p$} with \emph{$d$} Decision attributes, Algorithm type: \emph{optimizer over $\mathcal{S}$},
        Search configuration: \emph{brute force / iterative / both},
        Iteration count: \emph{$T$ iterations}, Children per node:
        \emph{tree branching factor $c$}, Simulation count: \emph{$R$ runs},
        Objective weights: \emph{trade-off parameters
        $(\alpha,\beta,\gamma)$} and the local and global objectives and cost
        functions.
        \label{fig:construct}
    ]{%
        \includegraphics[width=0.98\columnwidth]{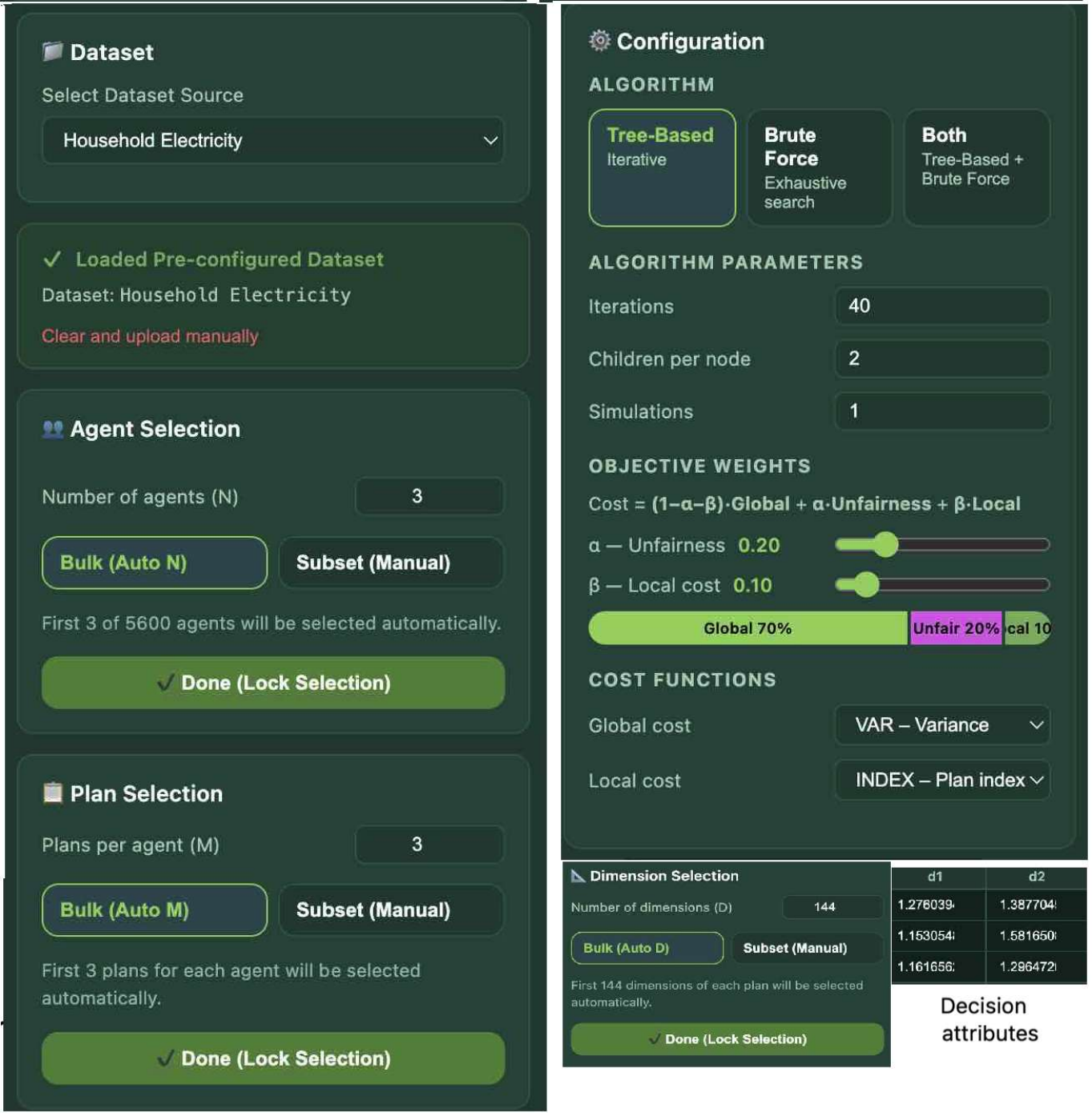}
    }

    \vspace{-1mm}

    \subfloat[
        Analyzer component of {\textsc{Argonaut}}, interface for interactive,
        iteration-by-iteration analysis of costs, convergence, plan changes,
        and comparative optimization outcomes.
        \label{fig:visualize}
    ]{%
        \includegraphics[width=0.98\columnwidth]{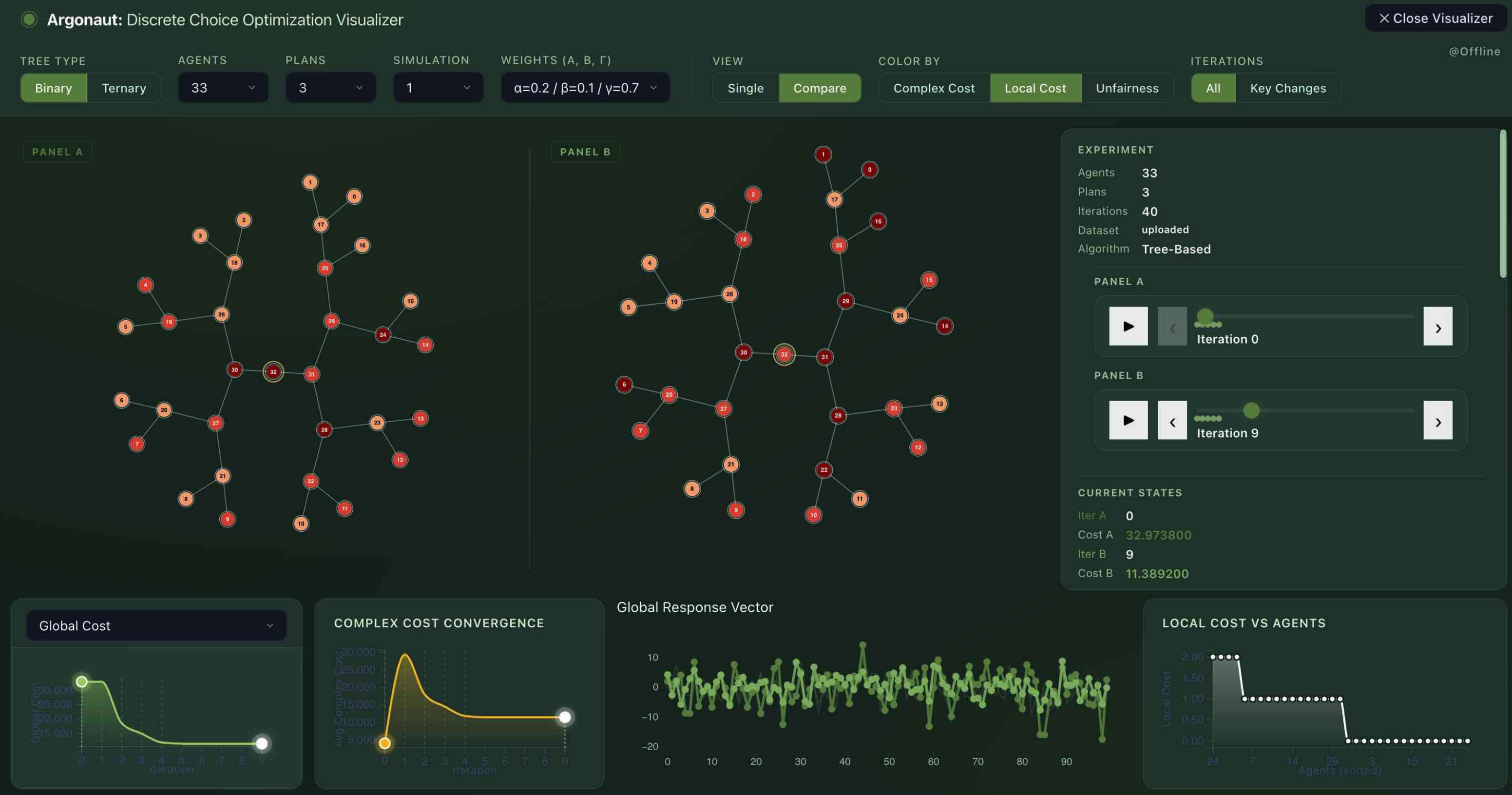}
    }

    \vspace{-1mm}
    \caption{Construct and analyzer components of {\textsc{Argonaut}}.}
    \label{fig:construct-analyze}
    \vspace{-2mm}
\end{figure}

\subsection{{\textsc{Argonaut}}}

Figure~\ref{fig:Design} shows the architecture of {\textsc{Argonaut}}. 

\noindent \textbf{Construct} (Figure~\ref{fig:construct}). Users can define, remove, and edit on the fly the number of agents $A={a_1,\ldots,a_n}$, their plans $P_i$, the decision attributes $d$, and the objectives and trade-off parameters that form the decision space $\mathcal{S}$.

\noindent \textbf{Optimize} (Figure~\ref{fig:Design}). {\textsc{Argonaut}} supports multiple algorithmic backends, such as brute-force search, brute-force search with pruning or centralized or decentralized iterative search with multiple algorithms available within each paradigm. A configurable input space is designed to support these multiple backends, combine and compare algorithms and also ranking iterative solutions against the global optimum from brute-force search. 


\noindent \textbf{Analyze} (Figure~\ref{fig:visualize}). {\textsc{Argonaut}} brings multiple interactive views of the optimization process into one place, covering cost, convergence, optimality, and solution comparison. These views show how global, local, and unfairness costs evolve, how plan selections change across iterations, and how local--global trade-offs are balanced, making the search process more transparent and interpretable.

\begin{figure}[h]
    \centering
    \includegraphics[width=0.50
    \textwidth]{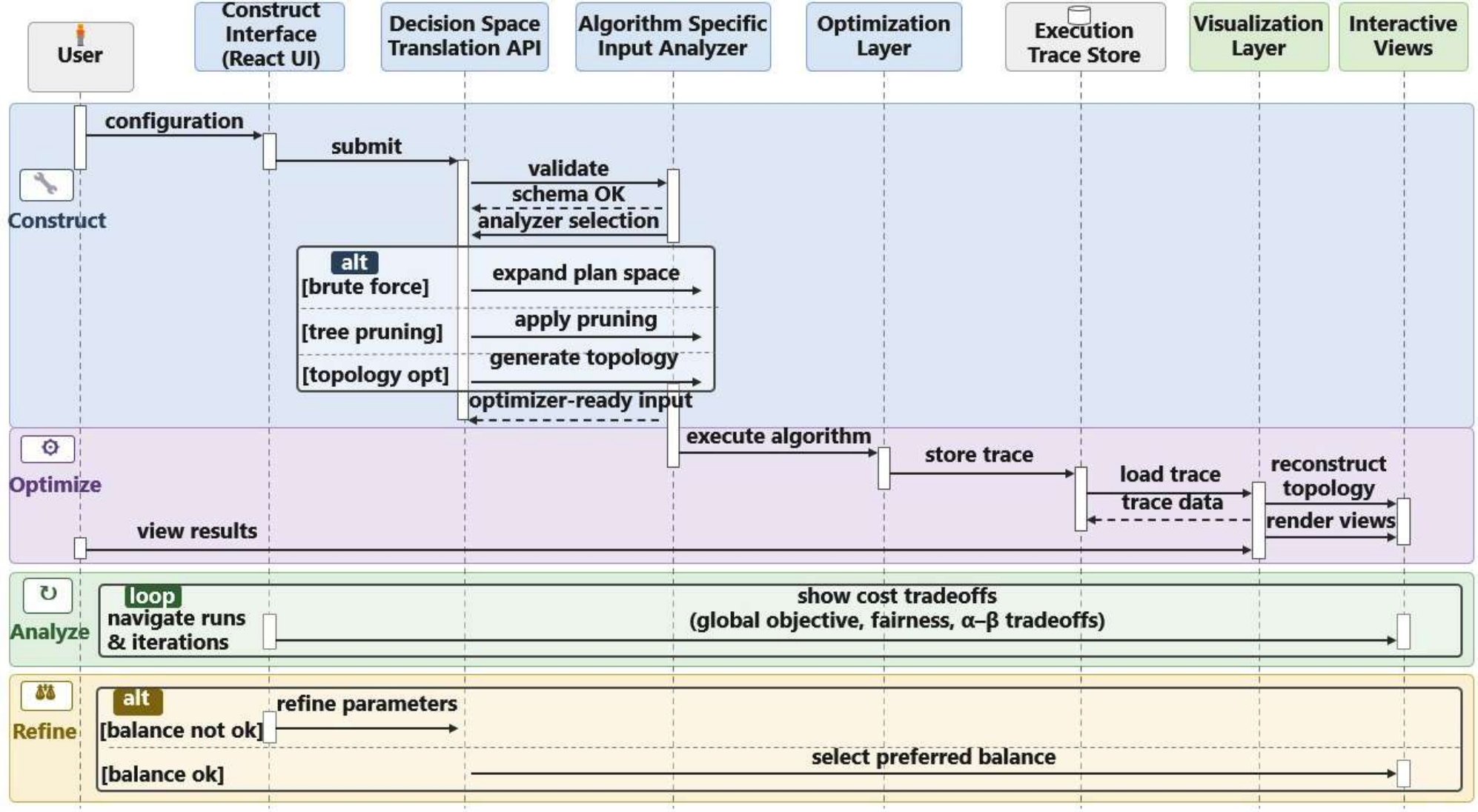}
  \caption{Decision space exploration sequence in {\textsc{Argonaut}}, showing how users configure  parameters, translate them into optimizer ready inputs, execute algorithms, inspect offline visual analytics, and iteratively refine the balance between global performance, fairness, and local cost.}
    \label{fig:sequence}
\end{figure}

This separation keeps {\textsc{Argonaut}} algorithm-agnostic and extensible while preserving a continuous \emph{construct--optimize--analyze} loop (full workflow illustrated in Figure~\ref{fig:sequence}).



\section{Implementation of {\textsc{Argonaut}}} \label{sec:implementation}

{\textsc{Argonaut}} is engineered around three goals: it should be \emph{lightweight}, \emph{highly portable}, and \emph{customizable}. 

{\em Construction Layer (Frontend \& Translation).} This layer is a React frontend on a Node.js stack~\cite{nodejs_about,react_components,vite_build}: React manages the interactive controls (topology, agents, plans, objective weights, cost and iteration views), Node.js serves assets with non-blocking I/O to stay {\em responsive} during optimization, and Vite~\cite{vite_build} compiles the frontend into static assets, keeping it \emph{lightweight}. A \emph{translation API}, a Python module behind a lightweight REST endpoint, then serializes the frontend state into a normalized JSON~\cite{nodejs_about} schema (agents, plans, constraints, objectives, trade-off weights, topology and algorithm) and converts it into the input format each algorithm requires, decoupling encoding so one decision space can drive different solvers, the key to \emph{customizable} design.

{\em Optimization Layer (Solver Backend).} In our design, algorithmic backends may be implemented in heterogeneous languages, each encapsulated by a language-specific solver wrapper exposing a uniform REST~\cite{fielding2000architectural} endpoint. Each wrapper derives from a common base via minimal customization, yielding a language-agnostic, portable, and extensible system.

{\em Visualization API (Rendering Engine).} The execution traces and topologies are converted into {\em interactive analytics} using Python NetworkX~\cite{networkx_docs}. NumPy computes layouts~\cite{numpy_docs}, Matplotlib renders reproducible figures~\cite{matplotlib_savefig}, and pandas processes  $.csv$ {\em logs} and tabular summaries~\cite{pandas_io}.

{\em Analysis Layer (Interactive Views)}: The analysis layer is supported by a lightweight Node.js interface that serves interactive views using dynamic tagged pointers for easy iteration-level analysis~\cite{nodejs_about,react_components,vite_build}.

{\em Deployment (Containerized Runtime).} {\textsc{Argonaut}} is containerized, packaging its runtime and dependencies into a single portable unit~\cite{bentaleb2022containerization} that is \emph{highly portable} and thus can be deployed both in the cloud and in standalone systems. A multistage Docker build separates the Node.js build environment from the production server, so the final container holds only the compiled front end and static resources~\cite{docker_multistage}, keeping it \emph{lightweight}. NGINX~\cite{docker_multistage} serves the interface and visualization output as static content through a thin production server~\cite{nginx_static}.


\begin{figure}[h]
    \centering
    \includegraphics[width=0.48\textwidth]{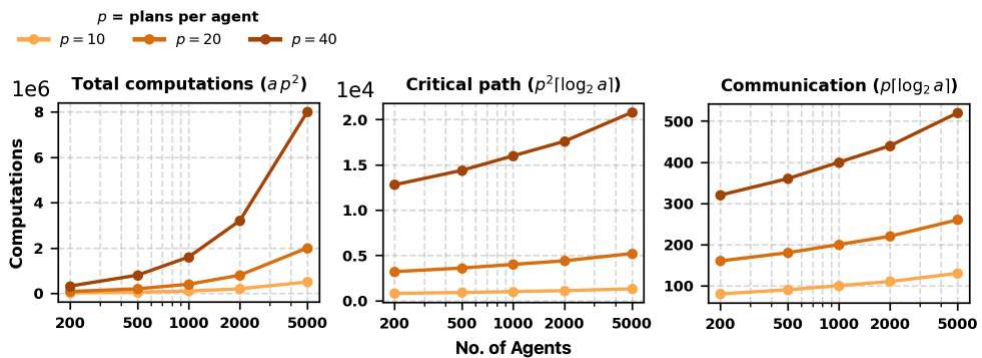}
  \caption{ \textbf{Computation load run in {\textsc{Argonaut}}, matching that reported for I-EPOS~\cite{pournaras2018decentralized}.} Scaling of total computations ($a\,p^{2}$), critical path ($p^{2}\lceil\log_{2}a\rceil$, the longest dependent chain of operations), and communication ($p\lceil\log_{2}a\rceil$) with the number of agents, for plans $p\in\{10,20,40\}$, run for 10 simulations and 40 iterations. Logarithmic tree depth keeps critical path and communication modest despite steep growth in total computation.}
\label{fig:complexity}
\end{figure}

\section{Evaluation} \label{sec:evaluation} 

For the proof-of-concept evaluation, we use synthetic and domain-specific discrete-choice datasets to assess the correctness, scalability, and effectiveness of {\textsc{Argonaut}}. The synthetic benchmark contains $1000$ agents, each with $16$ plans and $100$ decision attributes. Attribute values are drawn from a Gaussian distribution~\cite{wang2015fpga} and shaped as $y_{a,p,d}=x_{a,p,d}\cos(2\pi f_p d+\phi_a)$, giving each agent a distinct cosine-wave structure. This dataset is used to test both brute-force search, up to $1$M solutions, and decentralized iterative search. The real-world domain-specific datasets~\cite{Pournaras2019} cover energy management ($5600$ agents, $10$ plans, $144$ intervals), sensor data-sharing ($72$ agents, $3$ privacy-aware plans, $64$ attributes), and shared mobility ($2300$ agents, $2$--$50$ plans, $98$ bike stations).

\begin{figure}[h]
    \centering
    \includegraphics[width=0.48\textwidth]{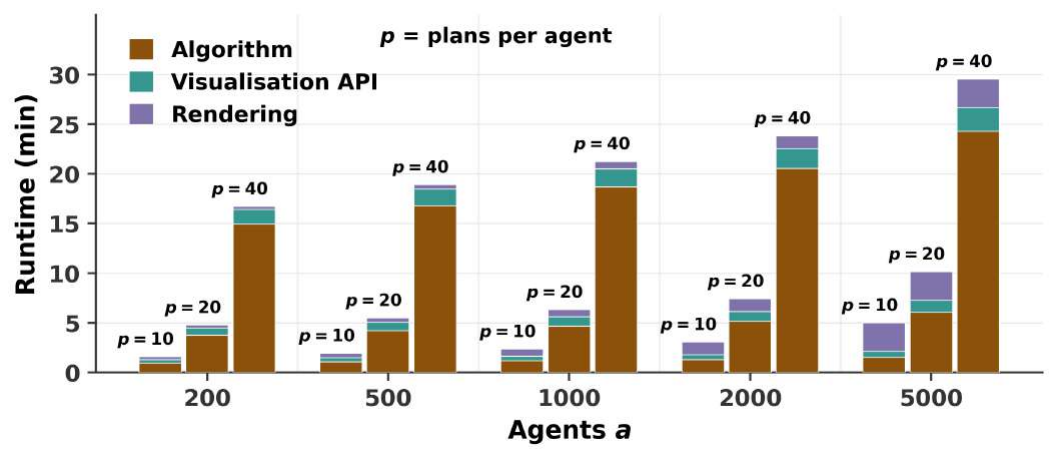}
\caption{\textbf{\textsc{Argonaut} keeps decentralized optimization within practical runtime limits (shown here for localhost), ranging from $26$ seconds to $28$ minutes  across up to $5000$ agents and $40$ plans per agent, while brute force already takes $110$ minutes for only $7$ agents and $6$ plans.} Agents: $a\in{200,500,1000,2000,5000}$, plans per agent: $p\in{10,20,40}$, with $100$ decision attributes. }
\label{fig:timing}
\end{figure}

We instantiate {\textsc{Argonaut}} with (i) a brute-force search that enumerates all feasible collective solutions to return the global optimum~\cite{nievergelt2000exhaustive} (ii) {\bf I-EPOS}, a collective learning algorithm~\cite{pournaras2018decentralized} for iterative decentralized optimization, as a representative backend. I-EPOS is chosen for its fully decentralized design and low computational and communication overhead, making it suitable for interactive visual exploration.


\begin{figure}[h]
    \centering
    \includegraphics[width=0.48\textwidth]{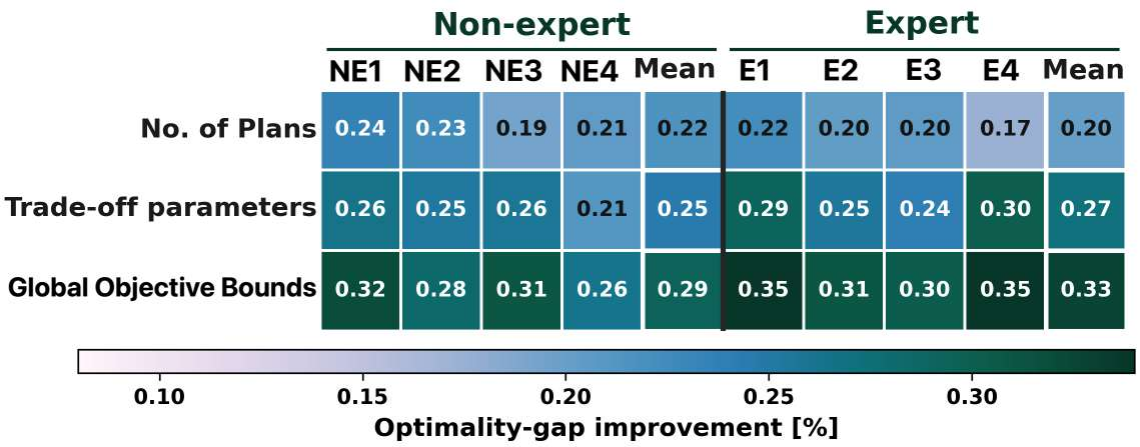}
\caption{\textbf{Optimality-gap improvement is highest when changing the target global objective bounds.} Optimality-gap improvement (\%) per configuration change across $4$ runs  for $4$ runs experts and $4$ non-expert users on $4$ datasets ($64$ runs per group). Cost-function and decision-attribute changes yield the largest improvement, while the number of plans and trade-off parameters have little effect, and the pattern holds across expertise. The optimality gap is $g=(C-C^{*})/C^{*}$, where $C$ is the achieved total cost and $C^{*}$ the optimal solution  from decentralized iterative search; improvement is the reduction in $g$ after the change.}
\label{fig:effect}
\end{figure}

\noindent\textbf{Correctness and Scalability}: {\textsc{Argonaut}} passes
correctness tests across both the centralized brute-force and decentralized
tree-based optimization modes. Component-level checks for data loading, plan parsing, objective computation, tree construction all pass successfully~\cite{barr2014oracle,rachatasumrit2012empirical}. Across $24$
independent runs, the solutions exported by {\textsc{Argonaut}} match
independently executed reference algorithms exactly, verified with a
similarity-checker script~\cite{noor2017programming}. This holds for the
brute-force mode over decision spaces up to $60$K solutions in cloud and 1M solutions in localhost machines, and for the decentralized tree-based
mode over $1000$ to $5600$ agents, $2$ to $50$ plans, and $64$ to $144$ decision
attributes, while varying the number of iterations, objectives,
topology structures, and trade-off parameters.

Having established correctness, we next examine how {\textsc{Argonaut}} scales, using the computation load it reports when running the I-EPOS tree-based decentralized optimizer~\cite{pournaras2018decentralized}. Figure~\ref{fig:complexity} decomposes it into total computation, the critical
path (the longest dependent chain of operations),
and communication cost. At $a=5000$ agents and $p=40$ plans, total computation
reaches $8\times10^{6}$ operations, while the critical path stays at
$2.08\times10^{4}$ and communication at $520$. Total work scales with agents and
plans, but coordination depth and communication stay orders of magnitude
smaller, matching the figures reported for I-EPOS. Figure~\ref{fig:timing} reports the measured runtime across configurations. The
tree-based optimization ranges from $28\,\text{s}$ at $(a,p)=(200,10)$ with  $100$ decision attributes to
roughly $28\,\text{min}$ at $(5000,40)$, increasing significantly with the
number of plans (pairwise $p=0.034$ for $10$ vs.\ $20$, $p=0.0056$ for $10$
vs.\ $40$, and $p=0.024$ for $20$ vs.\ $40$). The algorithm stage dominates,
on average $7.6\times$ longer than the visualization API and $7.4\times$
longer than rendering, with the latter two within a few percent of each other.  We conduct scalability testing for brute-force search across decision spaces of
increasing size: $5\times10^{4}$ collective solutions take approximately $110$
minutes, while $10^{6}$ solutions take up to $44$ hours. Its algorithm stage accounts for $\approx99\%$ of runtime,
running $111\times$ longer than the visualization API and $934\times$ longer
than rendering; these differences are significant ($p=0.0027$ algorithm vs.\
visualization, $p=0.00045$ visualization vs.\ rendering). On average, across
comparable configurations, brute force is roughly $\langle 85\rangle\times$
slower than the decentralized tree-based method, confirming its scalability
advantage.

{\em Localhost vs Cloud}: {\textsc{Argonaut}} is also cross-software compatible and the reported runtimes (Figure~\ref{fig:timing}) are measured on localhost across MacOS, Linux, and Windows (e.g. 4-core/8-thread CPU and 16 / 32 GB RAM). {\textsc{Argonaut}} is also available as a cloud-hosted service on Google Cloud Run using free credits; its runtime is on average approximately 40\% slower, as the minimal free-tier instance  ($512$~MiB memory, $0.08$~vCPU) shares resources with co-located workloads. Hosting it there tests the portability of the containerized design to the cloud, and adds flexibility for evaluation. Within Cloud Run, paid usage scales up to $8$~vCPU and $32$~GiB memory; in limited experiments this yields a $34\%$ reduction in visualization time and
$45\%$ in rendering time, and the containerized design ports directly to larger
instances on Compute Engine, GKE, or clouds such as AWS to recover
localhost-level performance. We will explore this in future for a production release on integration with transactional systems.


\noindent{\textbf{Effectiveness:}} To test the effectiveness of {\textsc{Argonaut}}, we evaluate it with four expert and four non-expert participants across four datasets: household electricity, sensor data sharing, shared mobility, and a synthetic Gaussian dataset. Each group performs $64$ runs ($4\times4\times4$) in total using decentralized iterative search, covering four datasets and four runs per dataset. In each run, a participant observes and interactively explores the optimization process in {\textsc{Argonaut}}, changes one control variable, either the trade-off parameters, the number of plans, or the target global objective bounds, and evaluates whether the edited decision space improves the solution rank by moving it closer to the global optimum, i.e., by reaching a lower-cost solution than the previous one. On average, across these $64$ runs for each group (Figure~\ref{fig:effect}), {\textsc{Argonaut}} reduces the optimality gap by up to $35.3\%$, with an average reduction of $24.87\%$, reaching the global optimum exactly in $12.56\%$ of cases. Changing the \emph{target global objective bounds} yields a statistically significant improvement over changing the number of plans ($p=0.0045$) and the trade-off parameters ($p=0.0003$), while the improvement from changing the number of plans over the trade-off parameters is not statistically significant ($p=0.073$). The mean difference between expert and non-expert users is also not statistically significant ($p=0.0894$), confirming that {\textsc{Argonaut}} makes effective optimization accessible to newcomers as well as experts. These results are based on 64 runs per group; a larger-scale user study is a direct and important extension we plan for future work.

\section{Conclusions} \label{sec:conclusions} 

We presented {\textsc{Argonaut}}, a lightweight, containerized, and highly customizable platform for interactive and visual exploration of distributed discrete-choice optimization. By supporting multiple algorithmic backends and uniting construction, optimization, and analysis in one interactive loop, {\textsc{Argonaut}} enables users to observe and navigate optimization runs within seconds to a few minutes, making it practical for prompt shared-resource management decisions in smart-city infrastructure. Guided by these views, users could improve solution optimality on average by $24.87\%$ across domains such as energy, mobility, transport, and logistics. Ongoing work extends {\textsc{Argonaut}} to additional backends, including decentralized iterative search under varying network topologies and brute-force search with pruning via Gurobi~\cite{nievergelt2000exhaustive}.

Building on this, our next step is to develop an explainability layer that surfaces \emph{why} the search behaves as it does, particularly by detecting and explaining when and why an agent changes its selected plan during convergence. This will move {\textsc{Argonaut}} from observable optimization toward fully explainable, human-in-the-loop decision-making, while also enabling the analysis of how such behaviours change in the presence of adversarial agents and how they can impact the overall optimization process.

\section*{Acknowledgment}
This work is funded by  UKRI Future Leaders Fellowship (MR\-/W009560\-/1): \emph{Digitally Assisted Collective Governance of Smart City Commons--ARTIO}' and UKRI Impact Acceleration Account grant (IAA4148): \emph{Collective Learning Optimization Algorithm}. The google cloud credits are funded by  UKRI AI Super Connector Fellowship from Imperial College London.
\bibliographystyle{IEEEtran}
\bibliography{acsos}

@misc{ortools,
  author       = {{Google}},
  title        = {{OR-Tools: Operations Research Tools}},
  howpublished = {\url{https://developers.google.com/optimization}},
  note         = {Accessed: 2026-06-18}
}

@misc{heuristiclab,
  author       = {{Heuristic and Evolutionary Algorithms Laboratory}},
  title        = {{HeuristicLab: An Environment for Heuristic and Evolutionary Optimization}},
  howpublished = {\url{https://github.com/heal-research/HeuristicLab}},
  note         = {Accessed: 2026-06-18}
}

@misc{iohanalyzer,
  author       = {Wang, Hao and Vermetten, Diederick and Ye, Furong and Doerr, Carola and B{\"a}ck, Thomas},
  title        = {{IOHanalyzer: Detailed Performance Analyses for Iterative Optimization Heuristics}},
  year         = {2020},
  howpublished = {\url{https://arxiv.org/abs/2007.03953}},
  note         = {Accessed: 2026-06-18}
}

@misc{visualizing_branch_and_bound,
  author       = {Ozalt{\i}n, Osman Y. and Hunsaker, Brady and Ralphs, Ted K.},
  title        = {{Visualizing Branch-and-Bound Algorithms}},
  year         = {2007},
  howpublished = {\url{https://optimization-online.org/2007/09/1785/}},
  note         = {Accessed: 2026-06-18}
}

@misc{apollo_choice_modelling,
  author       = {{Apollo Choice Modelling}},
  title        = {{Apollo: An R Package for the Estimation and Application of Choice Models}},
  howpublished = {\url{https://www.apollochoicemodelling.com/}},
  note         = {Accessed: 2026-06-18}
}

@misc{gurobi, author = {{Gurobi Optimization, LLC}}, title = {{Gurobi Optimizer}}, howpublished = {\url{https://www.gurobi.com/}}, note = {Accessed: 2026-06-18} }

@article{pournaras2018decentralized,
  title={Decentralized collective learning for self-managed sharing economies},
  author={Pournaras, Evangelos and Pilgerstorfer, Peter and Asikis, Thomas},
  journal={ACM Transactions on Autonomous and Adaptive Systems (TAAS)},
  volume={13},
  number={2},
  pages={1--33},
  year={2018},
  publisher={ACM New York, NY, USA}
}

@article{pournaras2016self,
  title={Self-repairable smart grids via online coordination of smart transformers},
  author={Pournaras, Evangelos and Espejo-Uribe, Jose},
  journal={IEEE Transactions on Industrial Informatics},
  volume={13},
  number={4},
  pages={1783--1793},
  year={2016},
  publisher={IEEE}
}

@inproceedings{hinrichs2013cohda,
  title={{COHDA}: A combinatorial optimization heuristic for distributed agents},
  author={Hinrichs, Christian and Lehnhoff, Sebastian and Sonnenschein, Michael},
  booktitle={International Conference on Agents and Artificial Intelligence},
  pages={23--39},
  year={2013},
  organization={Springer}
}

@article{yeoh2010bnb,
  title={{B}n{B}-{ADOPT}: An asynchronous branch-and-bound DCOP algorithm},
  author={Yeoh, William and Felner, Ariel and Koenig, Sven},
  journal={Journal of Artificial Intelligence Research},
  volume={38},
  pages={85--133},
  year={2010}
}

@incollection{mariel2021software,
  author    = {Mariel, Petr and Hoyos, David and Meyerhoff, J{\"u}rgen and Czajkowski, Miko{\l}aj and Dekker, Thijs and Glenk, Klaus and Jacobsen, Jette Bredahl and Liebe, Ulf and Olsen, S{\o}ren B{\o}ye and Sagebiel, Julian and Thiene, Mara},
  title     = {Software},
  booktitle = {Environmental Valuation with Discrete Choice Experiments},
  series    = {SpringerBriefs in Economics},
  pages     = {125--129},
  publisher = {Springer},
  address   = {Cham},
  year      = {2021},
  doi       = {10.1007/978-3-030-62669-3_9}
}

@inproceedings{jornod2015swarmviz,
  title={SwarmViz: An open-source visualization tool for Particle Swarm Optimization},
  author={Jornod, Guillaume and Di Mario, Ezequiel and Navarro, Inaki and Martinoli, Alcherio},
  booktitle={2015 IEEE congress on evolutionary computation (CEC)},
  pages={179--186},
  year={2015},
  organization={IEEE}
}

@inproceedings{he2016improved,
  title={An improved visualization approach in many-objective optimization},
  author={He, Zhenan and Yen, Gary G},
  booktitle={2016 IEEE congress on evolutionary computation (CEC)},
  pages={1618--1625},
  year={2016},
  organization={IEEE}
}

@inproceedings{lee2019visualizing,
  title={Visualizing swarm behavior with a particle density map},
  author={Lee, Hyeon-Chang and Kim, Yong-Hyuk},
  booktitle={Proceedings of the Genetic and Evolutionary Computation Conference Companion},
  pages={415--416},
  year={2019}
}

@article{thole2020design,
  title={Design space exploration and optimization using self-organizing maps},
  author={Thole, Sidhant Pravinkumar and Ramu, Palaniappan},
  journal={Structural and Multidisciplinary Optimization},
  volume={62},
  number={3},
  pages={1071--1088},
  year={2020},
  publisher={Springer}
}

@article{ramu2022survey,
  title={A survey of machine learning techniques in structural and multidisciplinary optimization},
  author={Ramu, Palaniappan and Thananjayan, Pugazhenthi and Acar, Erdem and Bayrak, Gamze and Park, Jeong Woo and Lee, Ikjin},
  journal={Structural and Multidisciplinary Optimization},
  volume={65},
  number={9},
  pages={266},
  year={2022},
  publisher={Springer}
}

@article{yadav2025handling,
  title={Handling objective preference and variable uncertainty in evolutionary multi-objective optimization},
  author={Yadav, Deepanshu and Ramu, Palaniappan and Deb, Kalyanmoy},
  journal={Swarm and Evolutionary Computation},
  volume={94},
  pages={101860},
  year={2025},
  publisher={Elsevier}
}

@misc{nodejs_about,author = {{OpenJS Foundation}},title = {{About Node.js}},howpublished = {\url{https://nodejs.org/en/about}},note = {Accessed: 2026-06-24}}

@misc{react_components,author = {{Meta Open Source}},title = {{React: Components and Props}},howpublished = {\url{https://www.w3schools.com/react/react_components.asp}},note = {Accessed: 2026-06-24}}

@misc{vite_build,
author = {{Vite}},
title = {{Building for Production}},
howpublished = {\url{https://vite.dev/guide/build}},
note = {Accessed: 2026-06-24}
}

@misc{docker_multistage,
author = {{Docker}},
title = {{Multi-stage Builds}},
howpublished = {\url{https://docs.docker.com/build/building/multi-stage/}},
note = {Accessed: 2026-06-24}
}

@misc{nginx_static,
author = {{NGINX}},
title = {{Serving Static Content}},
howpublished = {\url{https://docs.nginx.com/nginx/admin-guide/web-server/serving-static-content/}},
note = {Accessed: 2026-06-24}
}

@misc{networkx_docs,
author = {{NetworkX Developers}},
title = {{NetworkX Documentation}},
howpublished = {\url{https://networkx.org/documentation/stable/}},
note = {Accessed: 2026-06-24}
}

@misc{numpy_docs,
author = {{NumPy Developers}},
title = {{NumPy Documentation}},
howpublished = {\url{https://numpy.org/doc/stable/}},
note = {Accessed: 2026-06-24}
}

@misc{matplotlib_savefig,
author = {{Matplotlib Developers}},
title = {{matplotlib.pyplot.savefig}},
howpublished = {\url{https://matplotlib.org/stable/api/_as_gen/matplotlib.pyplot.savefig.html}},
note = {Accessed: 2026-06-24}
}

@misc{pandas_io,
author = {{pandas Developers}},
title = {{IO Tools}},
howpublished = {\url{https://pandas.pydata.org/docs/user_guide/io.html}},
note = {Accessed: 2026-06-24}
}

@article{wang2015fpga,
  title={{FPGA}-based design and implementation of arterial pulse wave generator using piecewise Gaussian--cosine fitting},
  author={Wang, Lu and Xu, Lisheng and Zhao, Dazhe and Yao, Yang and Song, Dan},
  journal={Computers in Biology and Medicine},
  volume={59},
  pages={142--151},
  year={2015},
  publisher={Elsevier}
}

@article{Pournaras2019,
author = "Evangelos Pournaras",
title = "{Agent-based Planning Portfolio}",
year = "2019",
month = "4",
url = "https://figshare.com/articles/dataset/Agent-based_Planning_Portfolio/7806548",
doi = "10.6084/m9.figshare.7806548.v6"
}

@inproceedings{nievergelt2000exhaustive,
  title={Exhaustive search, combinatorial optimization and enumeration: Exploring the potential of raw computing power},
  author={Nievergelt, J{\"u}rg},
  booktitle={International Conference on Current Trends in Theory and Practice of Computer Science},
  pages={18--35},
  year={2000},
  organization={Springer}
}

@article{barr2014oracle,
  title={The oracle problem in software testing: A survey},
  author={Barr, Earl T and Harman, Mark and McMinn, Phil and Shahbaz, Muzammil and Yoo, Shin},
  journal={IEEE transactions on software engineering},
  volume={41},
  number={5},
  pages={507--525},
  year={2014},
  publisher={IEEE}
}

@article{noor2017programming,
  title={Programming similarity checking system},
  author={Noor, Ahmad Shukri Mohd and Yunus, Farizah and Liang, Hoo Jian and Zin, Nur F Mat},
  journal={Journal of Telecommunication, Electronic and Computer Engineering (JTEC)},
  volume={9},
  number={3-5},
  pages={89--94},
  year={2017}
}

@inproceedings{rachatasumrit2012empirical,
  title={An empirical investigation into the impact of refactoring on regression testing},
  author={Rachatasumrit, Napol and Kim, Miryung},
  booktitle={2012 28th IEEE international conference on software maintenance (icsm)},
  pages={357--366},
  year={2012},
  organization={IEEE}
}

@inproceedings{colella2020human,
  title={Human strategic steering improves performance of interactive optimization},
  author={Colella, Fabio and Daee, Pedram and Jokinen, Jussi and Oulasvirta, Antti and Kaski, Samuel},
  booktitle={Proceedings of the 28th ACM conference on user modeling, adaptation and personalization},
  pages={293--297},
  year={2020}
}

@article{klau2010human,
  title={Human-guided search},
  author={Klau, Gunnar W and Lesh, Neal and Marks, Joe and Mitzenmacher, Michael},
  journal={Journal of Heuristics},
  volume={16},
  number={3},
  pages={289--310},
  year={2010},
  publisher={Springer}
}

@article{turan2024transition,
  title={Transition towards sustainable mobility: the role of transport optimization},
  author={Turan, Belma and Hemmelmayr, Vera and Larsen, Allan and Puchinger, Jakob},
  journal={Central European Journal of Operations Research},
  volume={32},
  number={2},
  pages={435--456},
  year={2024},
  publisher={Springer}
}

@article{blum2003metaheuristics,
  title={Metaheuristics in combinatorial optimization: Overview and conceptual comparison},
  author={Blum, Christian and Roli, Andrea},
  journal={ACM computing surveys (CSUR)},
  volume={35},
  number={3},
  pages={268--308},
  year={2003},
  publisher={Acm New York, NY, USA}
}

@incollection{kotthoff2016algorithm,
  title={Algorithm selection for combinatorial search problems: A survey},
  author={Kotthoff, Lars},
  booktitle={Data mining and constraint programming: Foundations of a cross-disciplinary approach},
  pages={149--190},
  year={2016},
  publisher={Springer}
}

@article{xin2020general,
  title={A general framework for decentralized optimization with first-order methods},
  author={Xin, Ran and Pu, Shi and Nedi{\'c}, Angelia and Khan, Usman A},
  journal={Proceedings of the IEEE},
  volume={108},
  number={11},
  pages={1869--1889},
  year={2020},
  publisher={IEEE}
}

@article{karunasingha2022root,
  title={Root mean square error or mean absolute error? Use their ratio as well},
  author={Karunasingha, Dulakshi Santhusitha Kumari},
  journal={Information Sciences},
  volume={585},
  pages={609--629},
  year={2022},
  publisher={Elsevier}
}

@article{western20099,
  title={9. variance function regressions for studying inequality},
  author={Western, Bruce and Bloome, Deirdre},
  journal={Sociological Methodology},
  volume={39},
  number={1},
  pages={293--326},
  year={2009},
  publisher={SAGE Publications Sage CA: Los Angeles, CA}
}

@article{bentaleb2022containerization,
  title={Containerization technologies: Taxonomies, applications and challenges},
  author={Bentaleb, Ouafa and Belloum, Adam SZ and Sebaa, Abderrazak and El-Maouhab, Aouaouche},
  journal={The Journal of Supercomputing},
  volume={78},
  number={1},
  pages={1144--1181},
  year={2022},
  publisher={Springer}
}

@book{fielding2000architectural,
  title={Architectural styles and the design of network-based software architectures},
  author={Fielding, Roy Thomas},
  year={2000},
  publisher={University of California, Irvine}
}

\end{document}